\documentclass[twocolumn,showpacs,preprintnumbers,amsmath,amssymb]{revtex4}
\usepackage{graphicx}% Include figure files
\usepackage{dcolumn}% Align table columns on decimal point
\usepackage{bm}% bold math

\begin{document}

%\preprint{CCU-th/xxx05}

\title{Superfluid stability in BEC-BCS crossover}% Force line breaks with \\

\author{C.-H. Pao}
\author{Shin-Tza Wu}
\affiliation{%
Department of Physics, National Chung Cheng University, Chiayi
621, Taiwan
}%

\author{S.-K. Yip}
% \homepage{http://www.Second.institution.edu/~Charlie.Author}
\affiliation{ Institute of Physics, Academia
Sinica, Nankang, Taipei 115, Taiwan}%

\date{June 13, 2005}
%\date{\today}% It is always \today, today,
             %  but any date may be explicitly specified

\begin{abstract}
We consider a dilute atomic gas of two species of fermions with
unequal concentrations under a Feshbach resonance. We find that
the system can have distinct properties due to the unbound
fermions. The uniform state is stable only when either (a) beyond
a critical coupling strength, where it is a gapless superfluid, or
(b) when the coupling strength is sufficiently weak, where it is a
normal Fermi gas mixture. Phase transition(s) must therefore occur
when the resonance is crossed.
\end{abstract}

     % PACS, the Physics and Astronomy
     % Classification Scheme.
%\keywords{Suggested keywords}%Use showkeys class option if keyword
                              %display desired
\maketitle

%\section{\label{sec:level1}First-level heading:\protect\\ The line
%break was forced \lowercase{via} \textbackslash\textbackslash}

%-------------------------------------

Feshbach resonance \cite{Feshbach} has opened up a new playground
for the field of cold trapped atoms.  Using this resonance, the
effective interaction between the atoms can be varied over a wide
range. In particular,  for two fermion species with a Feshbach
resonance between them, the ground state can be tuned from a
weak-coupling Bardeen Cooper Schrieffer (BCS) superfluid to a
strong coupling regime where the Fermions pair-up to form Bosons
which in turn undergo Bose Einstein Condensation (BEC)
\cite{EL,SRE93}.

Though this problem has been under intense theoretical \cite{theo}
and experimental \cite{expr} investigations, almost all works thus
far are restricted to the case where the concentrations of the two
fermionic species are equal.  We here generalize this study to the
case of unequal populations of the two species, and investigate in
detail the thermodynamic stability of this system, in particular
the question when the uniform state can be stable.

Studies of fermions with unequal populations or mismatched Fermi
surfaces and a pairing interaction have a long history. It was
studied by Fulde and Ferrell, Larkin and Ovchinnikov (FFLO)
\cite{FFLO} in the 1960's with relation to superconductivity in
materials with ferromagnetically coupled paramagnetic impurities.
It was found that in this case the system is likely to have an
inhomogeneous gapless superconducting phase. Advances in
techniques of manipulating dilute ultracold atoms
%\cite{BEC_rev}
have revived interests in the related problems \cite{FFLOn}.
 These studies, in our present language, are still restricted
to the weak-coupling regime.  We, however, would
extend our analysis to all coupling strengths.

In the ``canonical" problem of two species of fermions with equal
mass (say, spin up and spin down electrons) and equal
concentrations (thus a single Fermi surface), if the cross-species
interaction is varied from weak to strong coupling, at low
temperatures the system would undergo a {\em smooth crossover}
from a superfluid with loosely bound Cooper pairs (the ``BCS"
limit) to one with condensation of tightly bound bosonic molecules
(the ``BEC" limit) \cite{EL}. The situation, however, can be very
different if one considers two species of fermions with {\em
unequal concentrations} (i.e.~{\em mismatched} Fermi surfaces),
even if they have identical mass. This can be anticipated because,
on the one hand, far into the BCS side, the system is basically in
the FFLO regime \cite{FFLOn} and therefore must go into a
spatially inhomogeneous phase. On the other hand, in the far end
of the BEC side, the system is expected to behave like an ordinary
(weakly interacting) Bose-Fermi mixture and thus has a stable
homogeneous phase. Here the bosons are the fermion pairs and the
fermions are the ``leftover'' unpaired atoms of the majority
species. Deep into the BEC regime, the size of the Fermion pairs
is small and the interaction between the bosons and the leftover
unpaired fermions are expected to be weak. It is therefore a very
interesting question as to what happens in between. This is the
question we want to address in this paper.

For simplicity, we shall assume that the resonance is sufficiently
wide that the physics reduces effectively to a single channel
regime.  This is probably valid \cite{wide} for many Feshbach
resonances under current experimental investigations. Thus, in our
calculations, we would not invoke explicitly the presence of the
``closed channel'' which leads to this Feshbach resonance. We
simply model the fermions as interacting through a short-range,
s-wave effective interaction (dependent on the external magnetic
field) characterized by the corresponding scattering length $a$.
$1/a$ varies from $\infty$ for large negative detuning (closed
channel bound state energy much below continuum threshold) to
$-\infty$ for large positive detuning.

Now we proceed to the details of our calculation and results. We
consider two fermion species, denoted as ``spin'' $\uparrow$ and
$\downarrow$, of equal mass $m$.
 Because of the unequal concentrations of the two
species and the possible existence of pairing,
 it is useful to introduce three fields: the chemical
potentials $\mu_\sigma$ ($\sigma =  \uparrow$ or
$\downarrow$) and the pairing field $\Delta$.
We shall confine ourselves to zero temperature and generalize the BCS mean
field approach of \cite{SRE93}.
The excitation spectrum for each spin is
(see e.g. \cite{WY03} for details)
\begin{equation} E_\sigma ({\bf k})\, = \, \frac{
\xi_{\sigma}({\bf k}) - \xi_{- \sigma}({\bf k})}{ 2}\ +\
\sqrt{ \left ( \frac{ \xi_{\sigma}({\bf k}) + \xi_{-\sigma} ({\bf k})}
{2 } \right )^2\, + \, \Delta^2}\ ,
\end{equation}
where $\xi_{\sigma}({\bf k})\, =\, \hbar^2 k^2/2m - \mu_\sigma$ are
the quasi-particle excitation energies for normal fermions,
and $-\uparrow \equiv \downarrow$. The density of
each spin species is then
\begin{equation}
n_\sigma\ =\ \int \frac{ d^3 k }{ (2 \pi)^3} \left [ u_{\bf k}^2 f
(E_\sigma)\ +\ v^2_{\bf k} f( - E_{- \sigma}) \right ]\ ,
\label{ns}
\end{equation}
with the coherence factors ${\displaystyle u_{\bf k}^2 = 1- v_{\bf
k}^2\, =\, \frac{ E_\uparrow(k) + \xi_\downarrow(k) } {E_\uparrow
(k) + E_\downarrow (k) } }$. Here $f$ is the Fermi function.
 The equation for the order parameter $\Delta$ reads:
\begin{equation} - \frac{ m }{ 4 \pi a} \Delta\, =\, \Delta
 \int \frac{ d^3 k}{ (2\pi)^3}
\left [ \frac{ 1 - f(E_\uparrow) - f(E_\downarrow) }{
E_\uparrow + E_\downarrow }\, -\, \frac{ m }{ \hbar^2 k^2} \right ]\
. \label{eqgap}
\end{equation}
 We solve equations (\ref{ns}) and (\ref{eqgap})
self-consistently for fixed  total density $n \equiv n_\uparrow +
n_\downarrow$ and  density difference $n_d \equiv n_\uparrow -
n_\downarrow$.  We shall always take $\uparrow$ to be the
majority species so that $n_d \ge 0$.

It is convenient to introduce the average chemical potential $\mu
\equiv (\mu_\uparrow + \mu_\downarrow)/2$ and the difference $h
 \equiv\, (\mu_\uparrow - \mu_\downarrow)/2 \, \ge\, 0$.
Then we have
\begin{equation}
 E_{\uparrow,\downarrow} ({\bf k})\, = \,
\sqrt{ \xi({\bf k})^2 + \Delta^2} \mp h \ ,
\end{equation}
where $\xi({\bf k}) \equiv \hbar^2 k^2/2m - \mu$. Hence
$E_{\downarrow}(k) > 0$ always. From Eq. (\ref{ns}) we get
\begin{equation}
n_d = \ \int \frac{ d^3 k }{ (2 \pi)^3}  f (E_\uparrow({\bf k}))
\label{nd}
\end{equation}
and so the integration is only over the region where
$E_{\uparrow}({\bf k}) < 0$. In the following, it is useful to
note that the smallest (or most negative) $E_{\uparrow}({\bf k})$
occurs at $\xi({\bf k}) = 0$ for $\mu > 0$, where it is $\Delta -
h$, and at $k = 0$ for $\mu < 0$, where it is $\sqrt{ \mu^2 +
\Delta^2} - h$.

As in the case of equal concentrations, it is convenient to
express our results in dimensionless variables. We shall define an
inverse length scale $k_F$ through the total density $n$ via $k_F
\equiv (3 \pi^2 n)^{1/3}$, and an energy scale $\epsilon_F \equiv
\hbar^2 k_F^2 / 2 m$. We thus write $\tilde \mu \equiv
\mu/\epsilon_F$, $\tilde h \equiv h/\epsilon_F$, $\tilde \Delta
\equiv \Delta/\epsilon_F$, $\tilde n_d \equiv n_d/n$, and define
the dimensionless coupling constant $g \equiv 1/ (\pi k_F a)$,
which varies from $\infty$ for large negative detuning
%(closed channel bound state energy $\gg$ continuum threshold)
to $-\infty$ for large positive detuning.

We now describe the results of our calculations. We first make
contact with the BCS-BEC cross-over for equal concentrations. The
inset of Fig. 1 shows the typical behaviors of $\tilde \mu$,
$\tilde \Delta$ and $\tilde h$ as a function of $g$ for a given
density difference $\tilde n_d$.
 The behavior of $\tilde \mu$ or $\tilde \Delta$ is
similar to that in the case of equal concentrations  \cite{SRE93}.
For example, $\tilde \mu$ is large and negative in the BEC limit
whereas it is of order $1$ in the BCS regime. Unlike that case
however, $g$ has to be larger than a minimum coupling $g_c$ in
order for a finite order parameter $\Delta$ to exist. For $g <
g_c$, Eq. (\ref{eqgap}) requires that $\Delta = 0$ and the system
is in the normal state. (For clarity of this inset, we plot only
the $\Delta \ne 0$ solutions.)
 The main Fig. 1 shows $\tilde h$ as function of $g$ in the
intermediate regime ($|g| \lesssim 1$) for three different $\tilde
n_d$ (0.2, 0.5, and 0.8). The horizontal dotted lines indicate the
normal state in which the gap function $\Delta$ is zero (described
above).

The behavior for $g < g_c$ is easy to understand.
For sufficently large and negative $g$, the interaction is too weak
to produce pairing since the concentrations are unequal.
The system reduces to a Fermi gas.  In this case,
 the chemical potentials are given by
$\mu_\sigma = (6 \pi^2 n_\sigma)^{2/3}/(2m)$ which implies $\mu\,
=\, [(6 \pi^2)^{2/3}/4m](n_\uparrow^{2/3} + n_\downarrow^{2/3})$ and
$h\, =\, [(6 \pi^2)^{2/3}/4m)](n_\uparrow^{2/3} -
n_\downarrow^{2/3})$ (both independent of $g$). On the other hand, for
large and positive $g$ (the strong-coupling BEC
limit), one can show  \cite{Yip02}
 from Eqs. (\ref{ns}), (\ref{eqgap}) and (\ref{nd}) that
both $h$ and $|\mu|$ are large and to leading order given by
$\hbar^2/(2 m a^2)$.  However, $(\mu + h)/{2} $ $= \mu_{\uparrow}$
$= (6 \pi^2 n_d )^{ 2/3}/ (2m)$ $\ll |\mu|$ or $h$.  These
expressions simply reflect that the system becomes a Bose-Fermi
mixture with boson concentration $n_{\downarrow}$ and free fermion
concentration $n_d$.

Notice that the lines (for $\Delta \ne 0$) of $\tilde h$ versus
$g$  cross each other from small to large $\tilde n_d$ near $g
\sim 0.15$. For $g \gtrsim 0.17$, $\tilde h$ increases with
$\tilde n_d$ for fixed $g$. For $g \lesssim 0.15$, $\tilde h$
decreases as $\tilde n_d$ increases when the coupling strength is
fixed. We shall return to these features again below.

Now we make contact with the superconductivity literature. It is
helpful here to note that $h$ plays the role of an effective
external Zeeman field.  We plot in Fig. 2 $\tilde \Delta$ as a
function of $\tilde h$ for various coupling strengths $g$. The
horizontal portion of each curve corresponds to $n_d = 0$.  In
this region, $ h < \Delta$ and so that $E_{\uparrow}({\bf k}) > 0$
for all $\bf k$ (see Eq. (\ref{nd})). The other part of the curve
corresponds to $n_d > 0$, and exists only in the region  $h >
\Delta$ (More precisely, for larger $g$ where $\mu$ becomes
negative, this condition should read $h > \sqrt{\mu^2 +
\Delta^2}$). For small $g$ ($\lesssim 0.1$), $\tilde \Delta$
decreases with decreasing $\tilde h$. This solution is the
generalization of that first discovered by Sarma \cite{Sarma}. We
find that this superfluid state corresponds to one where $\tilde
n_d$ increases with decreasing $\tilde h$, and hence unstable (to
be discussed again below). For sufficiently large coupling ($g
\gtrsim 0.17$), $\tilde \Delta$ decreases with increasing $\tilde
h$.  This state has $\tilde n_d$ increases with $\tilde h$, and
satisfies one of the stability conditions to be discussed below.

In Fig. 3 the chemical potential difference $\tilde h$ is plotted as a
function of $\tilde n_d$. These results correspond to those
of Fig. 1 presented in a different manner.   Let us explain
how this graph should be read, with $g = -0.1$ as an example.
For $\tilde n_d = 0$, $\tilde h$ can take any value up to
 $\tilde h_1 \approx 0.5$ given
by the intersection of the line labelled by $g = -0.1$ with the
$\tilde n_d = 0$ axis.  This portion corresponds to the line with
$\Delta$ being a constant in Fig 2.  For $ 0 < \tilde n_d < 0.46$
the dependence of $\tilde h$ on $\tilde n_d$ is given by the solid
line labelled by $g = -0.1$.  This line corresponds to the state
with $\Delta \ne 0$ but $\Delta < h$ in Fig. 2 discussed above.
For $\tilde n_d > 0.46 $, the system enters into the normal state
with the $(\tilde h, \tilde n_d)$ relation represented by the
dotted lines, given by $\tilde h\, =\, 0.5 [ (1+ \tilde n_d)^{2
/3} - (1 -\tilde n_d)^{2 /3}]$ (see the discussion on Fig. 1
above). Lastly, for $\tilde n_d = 1$, $\tilde h$ can take any
value larger than $\tilde h_2 \equiv 0.5 \times 2^{2/3} \approx
0.79$. This is because this line corresponds simply to a Fermi gas
with only $\uparrow$ particles, and $h$ can take any value larger
than $\mu$ so that $\mu_{\downarrow} = (\mu - h)/2 < 0$. For $g
\gtrsim 0.1$, the graph can be read in a similar manner except
that the dotted line representing the normal state is not
involved.

For the uniform superfluid to be stable, two criterions
must be fulfilled \cite{WY03,Bedaque}.  First, the susceptibilities
matrix
$\partial n_{\sigma} / \partial \mu_{\sigma'}$ can have only
positive eigenvalues.  One can
show that this requires that $\partial \tilde n_d / \partial \tilde h$,
evaluated at constant $g$, must be positive \cite{matrix}.  That is,
the plot of $\tilde h$ versus $\tilde n_d$ must have positive
slope.
From Fig. 3, we see that for small $g$ ($ \lesssim 0.1$), the
slope of each curve is always negative which indicates the
superfluid state is unstable.  For $g$
greater than about 0.1, the slope of these curves change sign at some
$\widetilde{n}_d$  between $0$
and $1$. In this case a stable superfluid state may occur
for sufficiently large $\tilde n_d$. After $g \gtrsim 0.17$,
the slopes of these curves are positive for all $n_d \ge 0$
and the above stability criterion is satisfied for all
$n_d$.

The second stability criterion is that the superfluid
density $\rho_s$ must be positive \cite{WY03}.
$\rho_s$ can be evaluated as discussed in \cite{WY03}.
 The analytic result can be expressed
as
\begin{equation}
\frac{\rho_s}{n} =\
 \left [ 1- \frac{ \theta(\sqrt{ \mu^2+ \Delta^2} -h)
\tilde k_1^3 +
\tilde k_2^3 }{2 \sqrt{1- ( \Delta /h )^2} } \right ] \ ,
\label{rhos}
\end{equation}
with $\tilde k_{1,2} = [ (\tilde \mu \mp \sqrt{\tilde h^2- \tilde
\Delta^2})]^{1/2}$ and $\theta(x)$ is the step function. The line
$\rho_s = 0$ is plotted as the dashed lines in Fig. 3, with
$\rho_s < 0$ below and $\rho_s > 0$ above. Thus the states
correspond to $\Delta \ne 0$ below this dashed line are all
unstable.  $\rho_s < 0$ indicates that the system is unstable
towards a state with spatially varying $\Delta$ and hence a state
such as FFLO can be more preferable. From our results, it turns
out that the condition for $\rho_s > 0$ is actually a slightly
weaker requirement than the positive susceptibility discussed in
the last paragraph (though we are not aware of any reason why it
must be so).

We here note that, for $\tilde n_d \to 0$, the location where
$\rho_s$ changes sign is exactly at $\mu = 0$.  Though this can be
seen from Eq. (\ref{rhos}), it is more instructive to return to
the more basic equation for the superfluid density: $\rho_s = n +
\rho_p$.  Here $\rho_p$, the paramagnetic response, is related to
the backflow of quasiparticles and is given by \cite{WY03} $\rho_p
= - \frac{1}{6 \pi^2 m} \int_0^{\infty}
 dk k^4 \delta (E_{\uparrow}(k))$.
For $\mu < 0$, the solution to $E_{\uparrow}(k) = 0$ exists only
when $h > \sqrt{\mu^2 + \Delta^2}$ and takes place at $k = k_2$
with $k_2^2/2 m = \sqrt{h^2 - \Delta^2} + \mu$. For $n_d \to 0^+$,
$h$ is just slightly larger than $\sqrt{\mu^2 + \Delta^2}$ [see
Eq. (\ref{nd})]. $k_2$ is small and hence $\rho_p \to 0^+$ and
$\rho_s \approx n > 0$. For $\mu > 0$, $E_{\uparrow}(k) = 0$
happens when $h >  \Delta $ and take place at two $k$ values: $k =
k_2$ as already given in the $\mu < 0 $ case above and $k = k_1$
with $k_1^2/2 m = - \sqrt{h^2 - \Delta^2} + \mu$. For $n_d \to
0^+$, $h$ is just slightly larger than  $\Delta$ and the
$E_{\uparrow}(k) = 0$ points occur near $\xi(k) \approx 0$ hence
$\partial E_{\uparrow}(k) / \partial k \to 0$. Since $k_1$ and
$k_2$ are finite,  $\rho_p \to -\infty$ and $\rho_s < 0$.

Finally we show our phase diagram in Fig. 4 covering the entire
BEC to BCS regimes.  On the BEC side, with $g \gtrsim 0.17$, the
superfluid state is stable in which both the slope of $\tilde h
(\tilde n_d)$ and the superfluid density are positive (see Fig.
3). On the upper right of Fig. 4, the pairing order parameter
$\Delta$ is zero and the system is in the normal state. In
constructing the boundary of this phase, we have simply taken the
intersection of the full lines in Fig. 3 with the dotted lines.  A
more correct approach would involve the Maxwell construction.
However, to do this we also need to know the equation of state for
the non-uniform FFLO superfluid state beyond the weak-coupling
regime.
 Since this information is not yet available,
we shall leave this investigation to the future.

Lastly we discuss the ``breached gap'' state of Liu and Wilczek
\cite{LW}. For this state to exist, one need $E_{\uparrow} ({\bf
k}) < 0$ for a region of $k$ that lies between $k_1 < k <  k_2$
where $k_1 > 0$. ($k_1$, $k_2$ already given in the paragraph
before last).
 This is possible only if $E_{\uparrow}({\bf k})$ is not
monotonic with $k$ and hence $\mu > 0$.
Moreover, since $E_{\uparrow}({0}) > 0$,
we have $\sqrt{ \mu^2 + \Delta^2} > h$ yet $h > \Delta$.
Though there is a region of stability with $\mu > 0$
(the region near the upper right of Fig. 3), we find that
it rather corresponds to
 $\sqrt{ \mu^2 + \Delta^2} < h$.  (That this is the
case near $\tilde n_d \approx 1$ is obvious, since the line
$\tilde n_d = 1$ corresponds to $h > \mu$ and $\Delta \to 0$.
Note also on the dash-dotted lines where $\mu = 0$, we also have
$\sqrt{ \mu^2 + \Delta^2} < h$ since $h > \Delta$).  Therefore
gapless excitations exist only near $k = k_2$.  Moreover this
region has $n_{\uparrow} (k) = 1$ for $ 0 \le k \le k_2$ (and
$n_{\uparrow}(k) = n_{\downarrow}(k) = v^2(k) < 1$ for $k > k_2$),
thus the leftover unpaired majority spin-up particles form a
rather normal Fermi sphere with radius $k_2$, although the energy
required to create a hole $- E_{\uparrow}(k)$ can actually be
non-monotonic as a function of $k$, being maximum at an
intermediate value $k = \sqrt{ 2 m \mu}$ (where it is $h-\Delta$)
but not $k = 0$ (where it is $h - \sqrt{ \mu^2 + \Delta^2} $).

In conclusion, we have investigated the stability of a fermion
mixture with unequal concentrations under a Feshbach resonance. We
show that, in contrast to the case of equal concentrations, there
is {\it no} smooth BCS-BEC crossover.  The system is a uniform
stable superfluid Bose-Fermi mixture only for sufficiently large
coupling.  For weak interactions the normal state is the only
stable uniform state. The uniform state is unstable for
intermediate coupling strengths.  Phase transitions must occur
when the Feshbach resonance is varied between large positive
detuning and large negative detuning.

This research was supported by the National Science
Council of Taiwan under grant numbers
NSC93-2112-M-194-002 (CHP),
NSC93-2112-M-194-019 (STW) and NSC93-2112-M-001-016 (SKY),
with additional support from National Center for Theoretical
Sciences, Hsinchu, Taiwan.

%\vspace*{0.5in}

%\begin{acknowledgments}
%We wish to acknowledge the support of the author community in using
%REV\TeX{}, offering suggestions and encouragement, testing new versions,
%\dots.
%\end{acknowledgments}

%\newpage %Just because of unusual number of tables stacked at end
%\bibliography{apssamp}% Produces the bibliography via BibTeX.

\begin{thebibliography}{99}
\bibitem{matrix} We have evaluated the full susceptibility matrix
and checked that the eigenvalues are indeed all positive
when $\partial \tilde n_d / \partial \tilde h$ is positive.
\end{thebibliography}

\begin{figure}
\includegraphics{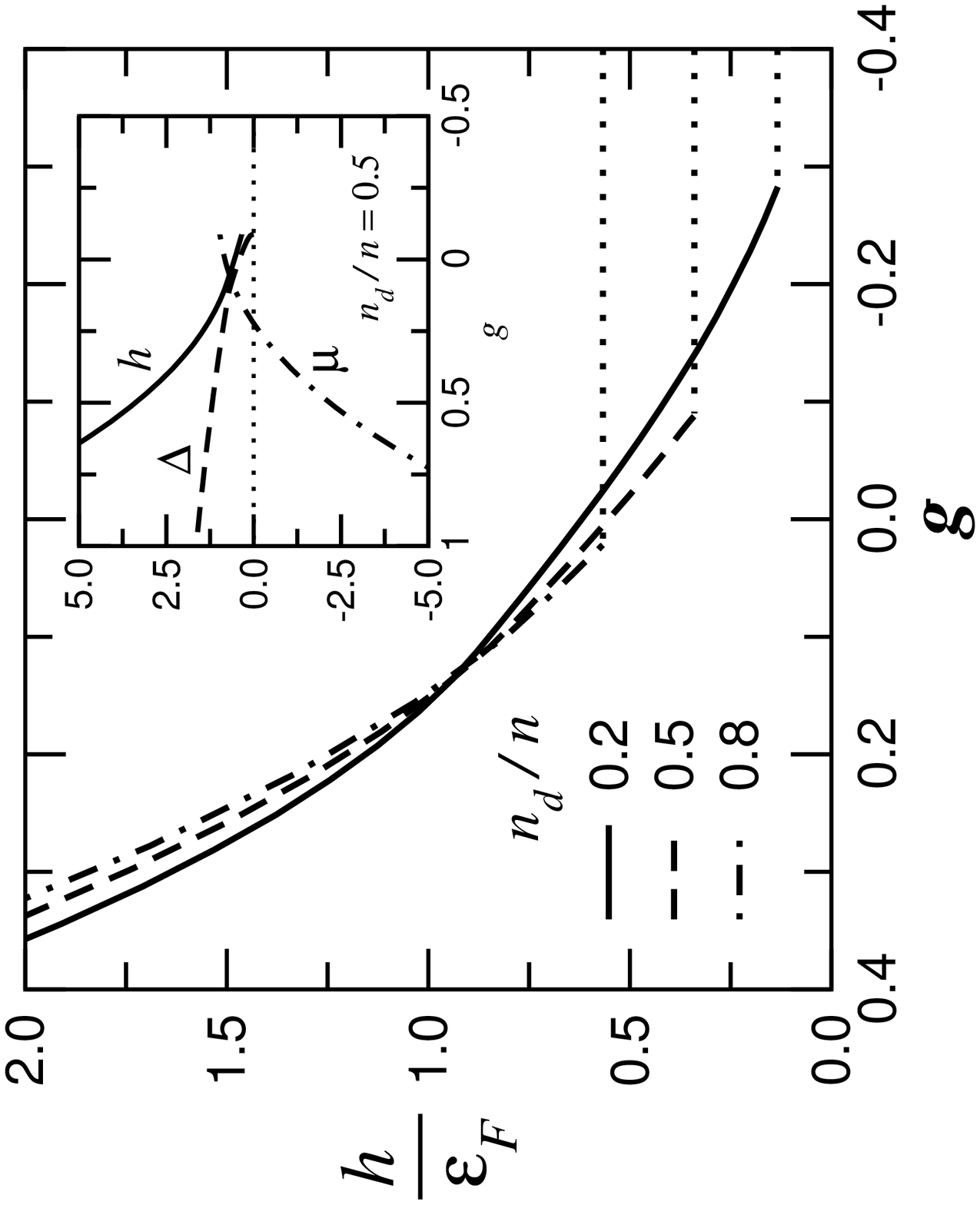} \vspace*{2.5in} \caption{Main figure: scaled
chemical potential difference $h/\varepsilon_F = \tilde h$ versus
the coupling constant $g$ for three different values of $n_d / n =
\tilde n_d$. Inset includes also the results for $\tilde \Delta$
and $\tilde \mu$ for $n_d / n = 0.5$.} \label{fig1}
\end{figure}

\begin{figure}
\includegraphics{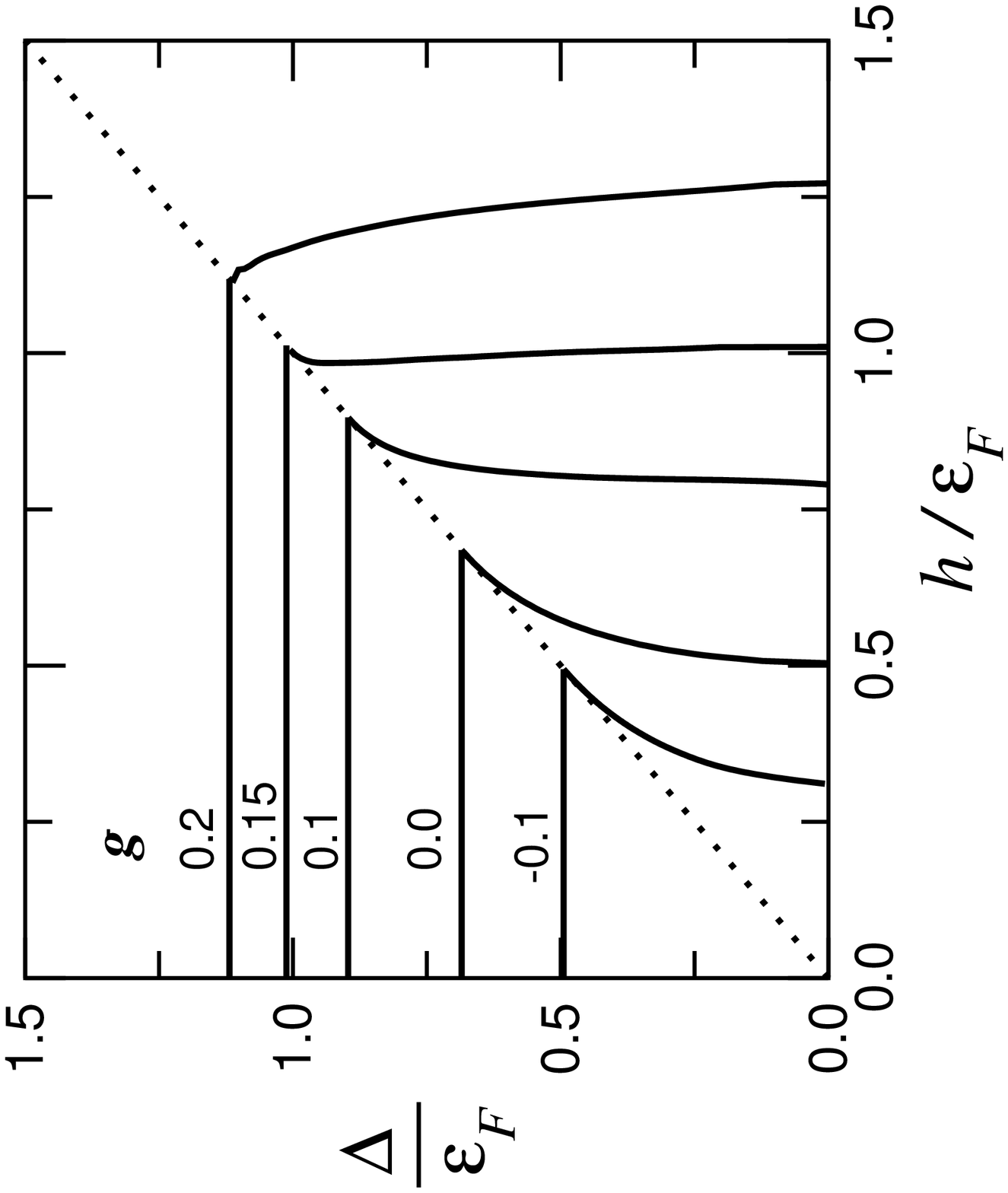} \vspace*{2.5in} \caption{ Scaled pair order
parameter $\Delta/\varepsilon_F = \tilde \Delta$ versus
$h/\varepsilon_F = \tilde h$ for given coupling constants $g$.}
\label{fig2}
\end{figure}

%\newpage
\begin{figure}
\includegraphics{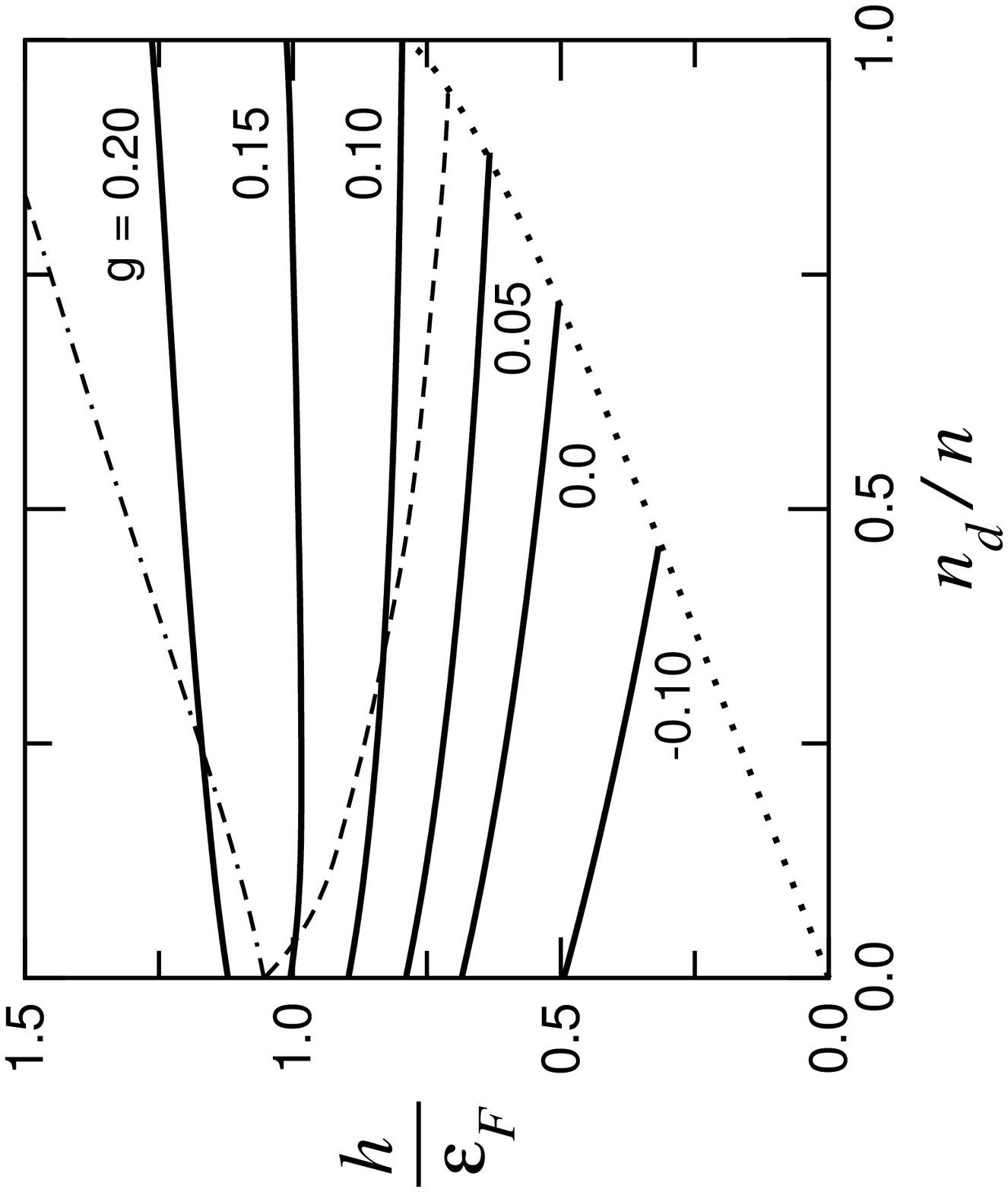} \vspace*{2.5in} \caption{ $h/\varepsilon_F =
\tilde h$ versus $n_d/n = \tilde n_d$ for constant $g$'s. Full
lines are for $\Delta \ne 0$ and the dotted lines are for $\Delta
= 0$.  The dashed line indicates where $\rho_s = 0$, with $\rho_s
> 0$ above it. The dot-dash line indicates $\mu = 0$, with $\mu <
0$ above this line.} \label{fig3}
\end{figure}

\begin{figure}
\includegraphics{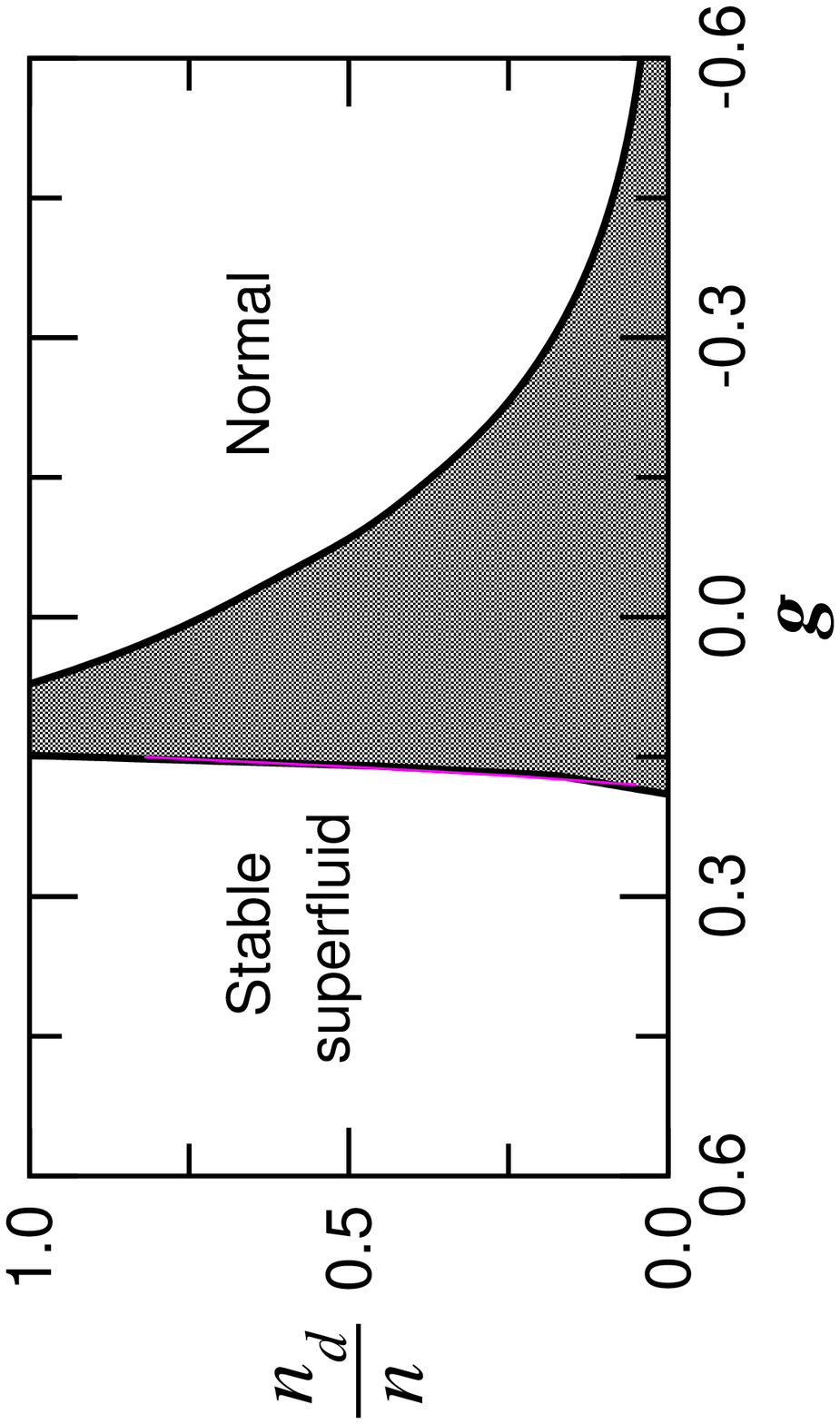} \vspace*{2.3in} \caption{ Phase diagram for
the two Fermion species with interspecies coupling $g$, with the
stable uniform phases (white regions) shown.  No uniform phases
are stable in the shaded region (except when $n_d/n = 0$ or $1$).
\label{fig4} }
\end{figure}

\end{document}